# The visualization of the space probability distribution for a moving particle I: in a single ring-shaped Coulomb potential

Yuan You[a)], Fa-Lin Lu [a)], Dong-Sheng Sun [a)], Chang-Yuan Chen [a)], and Shi-Hai Dong[b) 1]

[a)] New Energy and Electronic Engineering, Yancheng Teachers University, Yancheng, 224002, China

[b)] Laboratorio de Información Cuántica, CIDETEC, Instituto Politécnico Nacional,

Unidad Profesional Adolfo López Mateos, CDMX, C.P. 07700, Mexico

**Abstract**

We first present the exact solutions of the single ring-shaped Coulomb potential and then realize the visualizations of the space probability distribution for a moving particle within the framework of this potential. We illustrate the two-(contour) and three-dimensional (isosurface) visualizations for those specifically given quantum numbers ($n$, $l$, $m$) essentially related to those so-called *quasi* quantum numbers ($n'$, $l'$, $m'$) through changing the single ring-shaped Coulomb potential parameter $b$. We find that the space probability distributions (isosurface) of a moving particle for the special case $l = m$ and the usual case $l \neq m$ are spherical and circular ring-shaped, respectively by considering all variables $\vec{r} = (r, \theta, \varphi)$ in spherical coordinates. We also study the features of the relative probability values $P$ of the space probability distributions. As an illustration, by studying the special case of the quantum numbers ($n$, $l$, $m$)=(6, 5, 1) we notice that the space probability distribution for a moving particle will move towards two poles of axis $z$ as the relative probability value $P$ increases. Moreover, we discuss the series expansion of the *deformed* spherical harmonics through the orthogonal and complete spherical harmonics and find that the principal component decreases gradually and other components will increase as the potential parameter $b$ increases.

*Keywords*: Single ring-shaped Coulomb potential; space probability distribution; visualization; deformed spherical harmonics.

---

[1] Corresponding authors: Y. You (yuanyou_w@163.com); C.Y. Chen (yctcccy@163.net); S.H. Dong (dongsh2@yahoo.com).

# 1 Introduction

Since the ring-shaped non-central potentials have potential applications in quantum chemistry and nuclear physics, e.g., they might describe the molecular structure of Benzene and interaction between the deformed nucleuses, it is not surprising that the relevant investigations for them have attracted many attentions [1-20]. Based on previous study, we have known that this type of ring-shaped non-central potentials can be solved in spherical coordinates and also the system Hamiltonian with the hidden symmetry makes the bound state energy levels possess an "accidental" degeneracy, which arises from the SU(2) invariance of the Schrödinger Hamiltonian [1]. Generally speaking, the most popular ring-shaped non-central potentials are identified as the Coulomb or harmonic oscillator plus the ring-shaped part $1/(r^2 \sin^2\theta)$. In this work, we are concerned with only the single ring-shaped Coulomb potential in the limited space.

Many authors have obtained the radial and polar angular differential equations and also got their solutions in recent studies [7, 8, 13, 14, 21], but the whole space $\vec{r}=(r,\theta,\varphi)$ probability distributions of the moving particle in the single ring-shaped non-central fields have never been reported because of the complicated computation skills occurred in program. The main contributions mentioned above are either concerned with the radial part in the spherical shell $(r, r+dr)$ or with the angular parts in volume angle $d\Omega$ [22, 23]. This means that these studies are only related to one or two of three variables $(r,\theta,\varphi)$. To illustrate comprehensively the space probability distribution of the moving particle confined in the ring-shaped non-central Coulomb potential, the aim of this work is to realize their two- (contour) and three-dimensional (isosurface) visualizations by considering all variables. Such studies have never been done to our best knowledge.

The rest of this work is organized as follows. In Section 2 we first present the solutions of the studied quantum system and give the concrete expressions of the angular wave functions in order to compare with the usual spherical harmonics and also show their distinct properties. In Section 3 we make use of the calculation formula of the space probability distribution to illustrate their visualizations by overcoming the calculation skills in MATLAB program. In Section 4 we discuss the variation of the space probability distribution with the number of radial nodes, the variations with the relative probability value $P$ and those with the negative and positive ring-shaped Coulomb potential parameter $b$. The expansion coefficients of the deformed spherical harmonics are calculated in Section 5. Some concluding remarks are given in Section 6.

## 2. Exact solutions to single ring-shaped Coulomb potential

The single ring-shaped Coulomb potential is given by

$$V(r,\theta) = -\frac{Ze^2}{r} + \frac{\hbar^2}{2Mr^2}\frac{b}{\sin^2\theta}, \tag{1}$$

and the Schrödinger equation is written as

$$\left[-\frac{\hbar^2}{2M}\nabla^2 - \frac{Ze^2}{r} + \frac{\hbar^2}{2Mr^2}\frac{b}{\sin^2\theta}\right]\Psi(\vec{r}) = E\Psi(\vec{r}). \tag{2}$$

Take the wave function of the following form

$$\Psi(\vec{r}) = \frac{1}{\sqrt{2\pi}}\frac{u(r)}{r}H(\theta)e^{im\varphi}, \tag{3}$$

where $m = 0, \pm 1, \pm 2, \cdots$. Substitute (3) into (2) and get the respective radial and angular differential equations as

$$\frac{1}{\sin\theta}\frac{d}{d\theta}\left(\sin\theta\frac{dH(\theta)}{d\theta}\right) + \left(\lambda - \frac{m^2}{\sin^2\theta} - \frac{b}{\sin^2\theta}\right)H(\theta) = 0, \tag{4a}$$

$$\frac{d^2u(r)}{dr^2} + \left(\frac{2ME}{\hbar^2} + \frac{2M}{\hbar^2}\frac{Ze^2}{r} - \frac{\lambda}{r^2}\right)u(r) = 0. \tag{4b}$$

Take a new variable transform $x = \cos\theta$ and equation (4a) becomes

$$(1-x^2)\frac{d^2H(x)}{dx^2} - 2x\frac{dH(x)}{dx} + \left(\lambda - \frac{m^2+b}{1-x^2}\right)H(x) = 0. \tag{5}$$

Its solutions were given by the universal associated Legendre polynomials [22]

$$H_{l'm'}(x) = N_{l'm'} P_{l'}^{m'}(x) = N_{l'm'}(1-x^2)^{m'/2} \sum_{v=0}^{\left[\frac{l'-m'}{2}\right]} \frac{(-1)^v \Gamma(2l'-2v+1)}{2^{l'} v!(l'-m'-2v)! \Gamma(l'-v+1)} x^{l'-m'-2v}$$

$$= \sqrt{\frac{(2l'+1)\Gamma(l'+m'+1)}{2^{2m'+1}(l'-m')![\Gamma(m'+1)]^2}} (1-x^2)^{m'/2} {}_2F_1\left[-l'+m', l'+m'+1; m'+1; \frac{1-x}{2}\right].$$

(6)

where

$$m' = \sqrt{m^2+b} > 0, \quad l' = n_\theta + m', \quad \lambda = l'(l'+1), \quad n_\theta = 0, 1, 2, \cdots,$$

$$N_{l'm'} = \sqrt{\frac{(2l'+1)(l'-m')!}{2\Gamma(l'+m'+1)}}.$$

(7)

For $m' = -\sqrt{m^2+b} < 0$, notice that the differential Eq. (5) keeps invariant if replacing $m'$ by $-m'$, thus we know that the $P_{l'}^{-m'}(x)$ is also the solution of Eq. (5) and its definition is given by [24,25]

$$P_{l'}^{-m'}(x) = (-1)^m \frac{\Gamma(l'-m'+1)}{\Gamma(l'+m'+1)} P_{l'}^{m'}(x).$$

(8)

On the other hand, in terms of the spherical harmonics $Y_{lm}(\theta, \varphi)$ [26, 27]

$$Y_{lm}(\theta, \varphi) = (-1)^m \sqrt{\frac{2l+1}{4\pi} \frac{\Gamma(l-m+1)}{\Gamma(l+m+1)}} P_l^m(\cos\theta) e^{im\varphi},$$

(9)

We can define deformed spherical harmonics $Y_{l'm'}(\theta, \varphi)$

$$Y_{l'm'}(\theta, \varphi) = (-1)^m \sqrt{\frac{2l+1}{4\pi} \frac{\Gamma(l-m'+1)}{\Gamma(l+m'+1)}} P_l^{m'}(\cos\theta) e^{im\varphi},$$

(10)

with the following property

$$Y_{l'(-m')}(\theta, \varphi) = (-1)^m Y_{l'm'}(\theta, \varphi)^*.$$

(11)

Obviously, $m'$ is equal to $m$ when $b=0$, and Eqs. (8), (10), (11) will reduce to the results in central field. It is worth pointing out that all states $Y_{lm}(\theta, \varphi)$ are $2l+1$ for a certain value $l$, but states $Y_{l'm'}(\theta, \varphi)$ are 2 ($m \neq 0$), or 1 ($m=0$). This arises from the fact that the ring-shaped potential reduces the symmetry of the system and the degeneracy.

In Table 1 we list the analytical expressions of some given deformed spherical harmonics $Y_{l'm'}(\theta,\varphi)$ in the special cases of potential parameter $b=0$ and $b\neq 0$.

Table 1: Analytical expressions of the $Y_{l'm'}(\theta,\varphi)$ for special cases $b=0$ and $b\neq 0$

| l m | $Y_{lm}$ ( $b=0$ ) | $Y_{l'm'}$ ( $b\neq 0$ ) |
|---|---|---|
| 00 | $\dfrac{1}{\sqrt{4\pi}}$ | $\dfrac{1}{\pi}\left(2^{-1+\sqrt{b}}\Gamma(\tfrac{1}{2}+\sqrt{b})\sqrt{\dfrac{1+2\sqrt{b}}{\Gamma(1+2\sqrt{b})}}(\sin\theta)^{\sqrt{b}}\right)$ |
| 10 | $\sqrt{\dfrac{3}{4\pi}}\cos\theta$ | $\dfrac{1}{\pi}\left(2^{\sqrt{b}}\Gamma(\tfrac{3}{2}+\sqrt{b})\sqrt{\dfrac{3+2\sqrt{b}}{\Gamma(2+2\sqrt{b})}}\cos\theta(\sin\theta)^{\sqrt{b}}\right)$ |
| 1($\pm$1) | $\mp\sqrt{\dfrac{3}{8\pi}}\sin\theta e^{\pm i\varphi}$ | $\mp\dfrac{1}{\pi}\left(2^{-1+\sqrt{1+b}}\Gamma(\tfrac{1}{2}+\sqrt{1+b})\sqrt{\dfrac{1+2\sqrt{1+b}}{\Gamma(1+2\sqrt{1+b})}}(\sin\theta)^{\sqrt{1+b}}\right)e^{\pm i\varphi}$ |
| 20 | $\sqrt{\dfrac{15}{16\pi}}(3\cos^2\theta-1)$ | $\dfrac{1}{\pi}\left(2^{-\tfrac{1}{2}+\sqrt{b}}\Gamma(\tfrac{3}{2}+\sqrt{b})\sqrt{\dfrac{5+2\sqrt{b}}{\Gamma(3+2\sqrt{b})}}\left(-1+(3+2\sqrt{b})\cos^2\theta\right)(\sin\theta)^{\sqrt{b}}\right)$ |
| 2($\pm$1) | $\mp\sqrt{\dfrac{15}{8\pi}}\sin\theta\cos\theta e^{\pm i\varphi}$ | $\mp\dfrac{1}{\pi}\left(2^{\sqrt{1+b}}\Gamma(\tfrac{3}{2}+\sqrt{1+b})\sqrt{\dfrac{3+2\sqrt{1+b}}{\Gamma(2+2\sqrt{1+b})}}\cos\theta(\sin\theta)^{\sqrt{1+b}}\right)e^{\pm i\varphi}$ |
| 2($\pm$2) | $\sqrt{\dfrac{15}{32\pi}}\sin^2\theta e^{\pm i2\varphi}$ | $\dfrac{1}{\pi}\left(2^{-1+\sqrt{4+b}}\Gamma(\tfrac{1}{2}+\sqrt{4+b})\sqrt{\dfrac{1+2\sqrt{4+b}}{\Gamma(1+2\sqrt{4+b})}}(\sin\theta)^{\sqrt{4+b}}\right)e^{\pm i2\varphi}$ |
| 30 | $\sqrt{\dfrac{7}{16\pi}}(5\cos^3\theta-3\cos\theta)$ | $\dfrac{1}{\sqrt{3}\pi}\left(2^{\tfrac{1}{2}+\sqrt{b}}\Gamma(\tfrac{5}{2}+\sqrt{b})\sqrt{\dfrac{7+2\sqrt{b}}{\Gamma(4+2\sqrt{b})}}\cos\theta\left(-3+(5+2\sqrt{b})\cos^2\theta\right)(\sin\theta)^{\sqrt{b}}\right)$ |
| 3($\pm$1) | $\mp\sqrt{\dfrac{21}{64\pi}}(1-5\cos^2\theta)\sin\theta e^{\pm i\varphi}$ | $\mp\dfrac{1}{\pi}\left(2^{-\tfrac{1}{2}+\sqrt{1+b}}\Gamma(\tfrac{3}{2}+\sqrt{1+b})\sqrt{\dfrac{5+2\sqrt{1+b}}{\Gamma(3+2\sqrt{1+b})}}\left(-1+(3+2\sqrt{1+b})\cos^2\theta\right)(\sin\theta)^{\sqrt{1+b}}\right)e^{\pm i\varphi}$ |
| 3($\pm$2) | $\sqrt{\dfrac{105}{32\pi}}(\cos\theta-\cos^3\theta)e^{\pm i2\varphi}$ | $\dfrac{1}{\pi}\left(2^{\sqrt{4+b}}\Gamma(\tfrac{3}{2}+\sqrt{4+b})\sqrt{\dfrac{3+2\sqrt{4+b}}{\Gamma(2+2\sqrt{4+b})}}\cos\theta(\sin\theta)^{\sqrt{4+b}}\right)e^{\pm i2\varphi}$ |
| 3($\pm$3) | $\mp\sqrt{\dfrac{35}{64\pi}}\sin^3\theta e^{\pm i3\varphi}$ | $\mp\dfrac{1}{\pi}\left(2^{-1+\sqrt{9+b}}\Gamma(\tfrac{1}{2}+\sqrt{9+b})\sqrt{\dfrac{11+2\sqrt{9+b}}{\Gamma(1+2\sqrt{9+b})}}(\sin\theta)^{\sqrt{9+b}}\right)e^{\pm i3\varphi}$ |

Now, let us consider the radial differential equation. Substitute $\lambda = l'(l'+1)$ into (4b) and take $\chi = \tau r$, $s = 2MZe^2/\hbar^2$ and $\tau = (MZe^2/\hbar^2)\sqrt{-\hbar^2/2ME}$ then equation (4b) can be rewritten as

$$\frac{d^2 u(\chi)}{d\chi^2} + \left(\frac{s}{\chi} - \frac{1}{4} - \frac{l'(l'+1)}{\chi^2}\right)u(\chi) = 0. \tag{12}$$

Its solutions are nothing but the Coulomb case, i.e. [28]

$$u_{n'l'}(r) = \frac{1}{\Gamma(2l'+2)}\left[\frac{Z}{a_0}\frac{\Gamma(n'+l'+1)}{n_r!(n')^2}\right]^{1/2}\left(\frac{2Zr}{a_0 n'}\right)^{l'+1} e^{-\frac{Zr}{a_0 n'}} F\left(-n_r, 2l'+2, \frac{2Zr}{a_0 n'}\right), \tag{13}$$

where $n' = n_r + l' + 1 = n_r + n_\theta + m' + 1$, $n_r = 0, 1, 2, \cdots$ and $a_0 = \hbar^2/Me^2$ is the Bohr radius. Thus, the wave function in the whole space is written as

$$\Psi_{n'l'm}(\vec{r}) = \frac{1}{\sqrt{2\pi}}\frac{u_{n'l'}(r)}{r}H_{l'm'}(\cos\theta)e^{im\varphi}. \tag{14}$$

## 3. Isosurface and contour visualizations of space probability distributions

It is well known that the space probability distributions for a moving particle at the position $\vec{r} = (r, \theta, \varphi)$ can be calculated by

$$\rho = |\Psi_{n'l'm'}(\vec{r})|^2 = \frac{1}{2\pi}\frac{u_{n'l'}^2(r)}{r^2}H_{l'm'}^2(\cos\theta). \tag{15}$$

Obviously, this formula is independent of the azimuth angle $\varphi$ and thus symmetric with respect to the axis *z*. In order to display the space probability distribution we will transform (15) from original spherical coordinates to the popular Cartesian coordinates through the coordinate transformations $r = \sqrt{x^2 + y^2 + z^2}$, $\cos\theta = z/r$ and then obtain the corresponding space probability distribution $\rho(x, y, z)$.

For a given and definite space, we take a series of discrete positions and calculate their respective probability distribution values to realize their numerical calculations.

To improve the graphic resolution, we take *N* concrete positions in the whole space ($x, y, z$) and calculate density block, say *den*(*N*, *N*, *N*), which is composed of all values $w_{n'l'm'}$ for all $N \times N \times N$ positions. In this work, we take *N*=81. For given *quasi* quantum numbers $(n', l', m')$, we realize their isosurface (three dimensional) and contour (two dimensional) visualizations of the space probability distributions for different states $(n \leq 5)$ by using MATLAB program (see Tables 2 and 3).

## 4. Discussions

### 4.1 *Variation of space probability distribution with respect to the numbers of radial nodes*

In Table 2 we display the space probability distributions for three different cases $b = 0$ corresponding to the Coulomb potential, $b = 0.5$ and $b = 10$ corresponding to ring-shaped potentials. The unit in axis is taken as the Bohr radial $a_0$. To clearly visualize the internal structure of the graphics, we generate a section plane without considering those numerical values in the regions $x < 0, y < 0, z > 0$.

It is found that the graphics becomes compressed, i.e., the space probability distributions elongate along with the axis *x* and *y* and also the hole formed in the ring-shaped potentials expand towards outside as the ring-shaped potential parameter *b* increases. This can be well understood by the relations given in Eq. (7). We know that the value of the *quasi* quantum number *m'* increases relatively for a given quantum number *m*. When $l = m$, the isosurface of the density distribution is spherical, but for the case $l \neq m$, its isosurface is circularly ring-shaped.

In Table 3, the space probability distribution is projected to plane *yoz* and is shown to be symmetric with respect to the axis *y* and *z*. Here we only plot the graphics in the first quadrant through magnifying proportionally the space probability $|\Psi_{n'l'm'}(\vec{r})|^2$ and making it the maximum value to be 100, while the interval is taken as

10. There exists a corresponding balance among the density distributions in axis directions *x*, *y* and *z* since the sum of density distributions is equal to unit according to the normalization condition.

## 4.2 *Variation of the space probability distributions with respect to different relative probability values P*

In order to display the isosurface of the space probability distributions for different chosen relative probability values $P \in (0,100)\%$, we take the quantum numbers (*n*, *l*, *m*)=(6, 5, 1) as a typical example in Table 4. We find that for the smaller *P* the particle is distributed to almost all position spaces but for the larger *P*, we notice that the particle moves towards to two poles of axis *z*.

## 4.3 *Variations of the space probability distribution with respect to different potential parameter b*

For fixed quantum numbers $m$ and $n_\theta$, it is known from Eq. (7) that the *quasi* quantum number $m'$ becomes larger with the increasing *b* and thus leads to the increment of the *quasi* quantum number $l'$. In Table 5 we plot the space probability distributions of the quantum state $(n,l,m) = (4,0,0)$ for the case $b > 0$. Due to the change of *quasi* quantum number $m'$ which is arose from the potential parameter *b*, the number of radial nodes is changed accordingly. This leads to the extensions of the space probability distributions along with the axis *x* and *y* when the parameter *b* increases.

In Table 6, we give the comparison between the special cases for $b < 0$ and $b > 0$. Obviously, we find that the space probability distributions for the negative $b = -0.5$ in comparison with the special case $b=0$ are shrunk towards to the origin of the graphics. On the contrary, those for the positive case $b=0.5$ are enlarged towards outside. This kind of phenomenon can be understood well from analyzing the contributions of the potential parameter *b* to the Coulomb part. The original attractive

Coulomb potential becoming bigger or smaller relatively depends on the choice of the negative or positive *b*. As a result, the attractive force acting on the moving particle becomes larger or smaller. Naturally, this leads to the space probability distributions shrunk or extended.

## 5. Expansion coefficients of the deformed spherical harmonics

Since the spherical harmonics (9) are orthogonal and complete, then the deformed spherical harmonics (10) can be expanded by the usual spherical harmonics, i.e.

$$Y_{l'm'}(\theta,\varphi) = \sum_{l,m} a_{lm} Y_{lm}(\theta,\varphi), \tag{16}$$

where the expansion coefficients can be calculated by

$$a_{lm} = \sqrt{\frac{(2l'+1)(2l+1)(l-m)!\Gamma(l'-m'+1)}{4(l+m)!\Gamma(l'+m'+1)}} \int_{-1}^{+1} P_l^m(x) P_{l'}^{m'}(x) dx. \tag{17}$$

In Fig. 1 we plot the variation of the expansion coefficients with different parameter b where we take $m=1$, $l'-m'=2$, $b=0.5, 5, 10$. For $m \neq 1$, according to the orthogonal normalization condition in the $\varphi$ direction, each value of $a_{lm}$ is zero. For $m=1$, each value of $a_{lm}$ is also equal to zero for the case of odd value $l-m$. Thus, in Fig.1 we only present the case of even value $l-m$ for $m=1$. Notice that $P_3^1(x)$ is always biggest, but the principal component $P_3^1(x)$ decreases gradually, and other components will increase as the parameter *b* increases.

## 6. Concluding remarks

In this work we first presented the exact solutions to the single ring-shaped Coulomb potential and then realized the visualization of the space probability distributions for a moving particle within the framework of this potential. We have illustrated the two-(contour) and three-dimensional (isosurface) visualizations for some given *quasi* quantum numbers $(n',l',m')$ by taking different ring-shaped potential parameter *b*.

We have found that the space probability distributions of the moving particle in the cases of the $l=m$ and $l \neq m$ are spherical and circularly ring-shaped, respectively.

Moreover, we have studied the features of the relative values *P* of the space probability distributions. As an illustration, we have discussed the special case, i.e., the quantum numbers (*n, l, m*)=(6, 5, 1) and noticed that the space probability distributions for the moving particle will move towards two poles of axis *z* as the relative values *P* increase. We have also studied the series expansion of the deformed spherical harmonics by using the orthogonal and complete spherical harmonics and found that the principal component decreases gradually and other components will increase as the potential parameter *b* increases.

**Competing Interests**

The authors declare that there is no conflict of interests regarding the publication of this paper.

**Acknowledgments**: We would like to thank the kind referees for invaluable and positive suggestions which improved the manuscript greatly. This work is supported by the National Natural Science Foundation of China under Grant No. 11275165 and partially by 20170938-SIP-IPN, Mexico. Prof. You also acknowledges Jiangsu Overseas research & Training Program for University Prominent Young & Middle-aged Teachers and Presidents to support.

# References

[1] Quesne C 1988 J. Phys. A: Math. Gen. 21 3093

[2] Zhedanov A S 1993 J. Phys. A: Math. Gen. **26** 4633

[3] Chen C Y and Sun D S 2001 Acta Photonica Sin. **30** 104 (in Chinese)

[4] Sun D S and Chen C Y 2001 Acta Photonica Sin. **30** 539 (in Chinese)

[5] Dong S H, Sun G H and Lozada-Cassou M 2004 Phys. Lett. A **328** 299

[6] Guo J Y, Han J C and Wang R D 2006 Phys. Lett. A **353** 378

[7] Hartmann H 1972 Theor. Chim. Acta **24** 201

[8] Hartmann H and Schuck R 1980 Int. J. Quantum Chem. **18** 125

[9] Mandal B P 2000 Int. J. Mod. Phys. A **15** 1225

[10] Blado G G 1996 Int. J. Quantum Chem. **58** 431

[11] Gerry C C 1986 Phys. Lett. A **118** 445

[12] Chetouani L, Guechi L and Hammann T F 1992 J. Math. Phys. **33** 3410

[13] Chen C Y, Lu F L, Sun D S, and Dong S H 2013 Chin. Phys. B **22** 100302

[14] Chen C Y, Sun D S and Liu C L 2003 Phys. Lett. A **317** 80

[15] Sobhani H, Hassanabadi H 2016 Phys. Lett. B **760 1**

[16] Hassanabadi H, Kamali M, Molaee Z and Zarrinkamar S 2014 Chin. Phys. C **38** (3) 033102

[17] Chabab M, El Batoul A, Oulne, M, Hassanabadi H, Zare S. 2016 J. Korean Physical Society **69** (11) 1619

[18] Zarrinkamar S, Jahankohan K, Hassanabadi H 2015 Can. J. Phys. **93** (12) 1638

[19] Ikot A N, Obong, H P, Owate, I O, Onyeaju, M C, Hassanabadi, H 2015 Advances in High Energy Physics **2015** 632603

[20] Li W, Chen C Y, Dong S H 2017 Advances in High Energy Physics 2017 ID 7374256

[21] Chen C Y, Lu F L, Sun D S, You Y, and Dong S H 2015 Appl. Math. Lett. **40** 90

[22] Sun D S, You Y, Lu F L, Chen C Y, and Dong S H 2014 Phys. Scr. **89** 045002

[23] Sari R A, Suparmi A, and Cari C 2016 Chin. Phy. B **25** 010301

[24] Gradshteyn I S, and Ryzhik I M, Tables of Integrals, Series, and Products, sixth ed., Academic Press: Oxford, New York, 2000.

[25] Andrews L C, Special Functions of Mathematics for Engineers, second ed., Oxford University Press: Oxford, New York, 1998.

[26] Condon E U, Shortley G H, The Theory of Atomic Spectra, Cambridge University press: The Macmillan Company, New York, 1935.

[27] Strange P, Relativistic Quantum Mechanics, Oxford University Press: Oxford, New York, 1998.

[28] Landau L D and Lifshitz E M, Quantum Mechanics (Non-Relativistic Theory), 3rd ed. Pergamon Press: Oxford, New York, 1977.

Table 2: Three dimensional (isosurface) space probability distributions with a section plane

| $n$ | $l$ | $m$ | $B$ | | |
|---|---|---|---|---|---|
| | | | 0 | 0.5 | 10 |
| 1 | 0 | 0 | 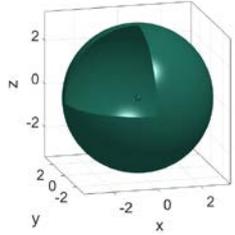 | 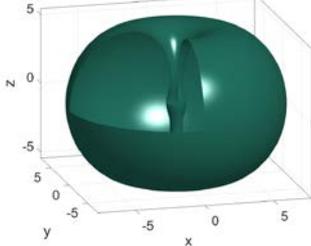 | 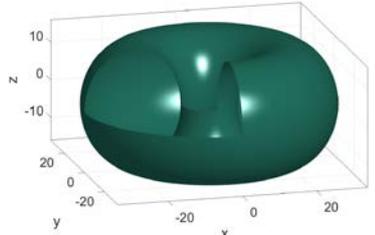 |
| 2 | 0 | 0 | 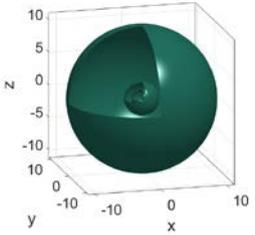 | 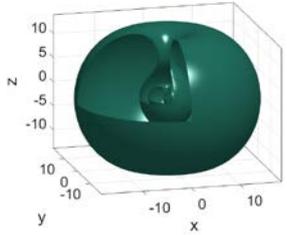 | 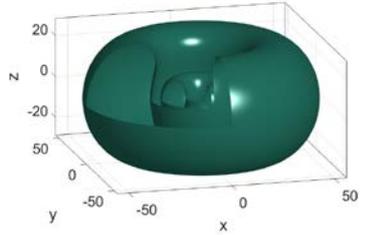 |
| 2 | 1 | 0 | 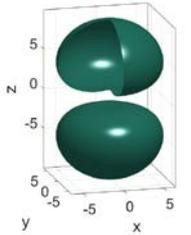 | 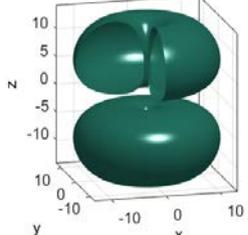 | 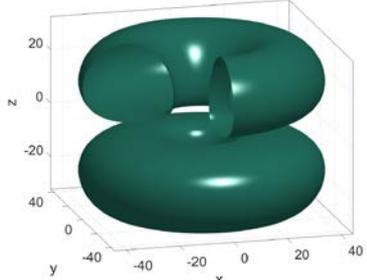 |
| 2 | 1 | 1 | 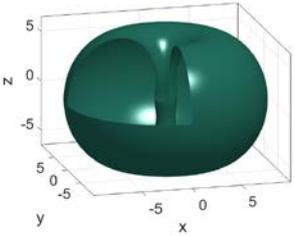 | 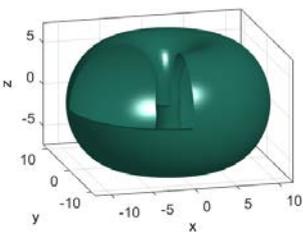 | 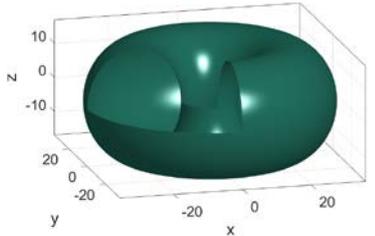 |
| 3 | 0 | 0 | 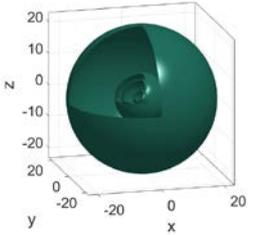 | 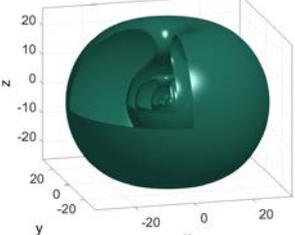 | 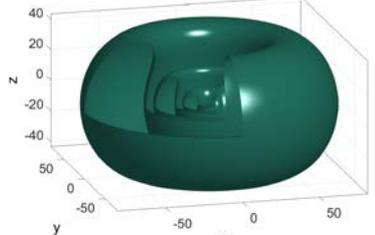 |

| 3 | 1 | 0 | 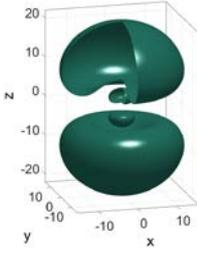 | 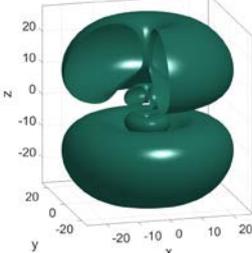 | 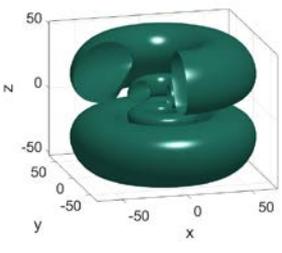 |
|---|---|---|---|---|---|
| 3 | 1 | 1 | 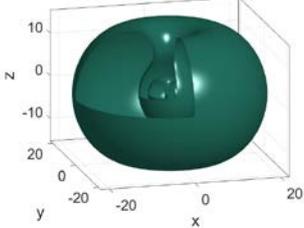 | 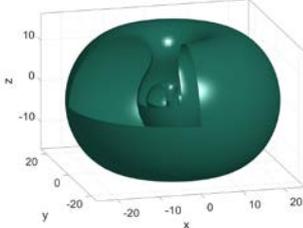 | 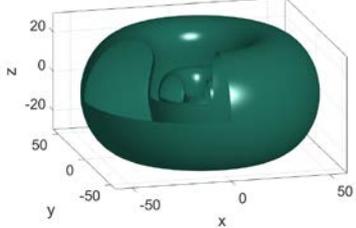 |
| 3 | 2 | 0 | 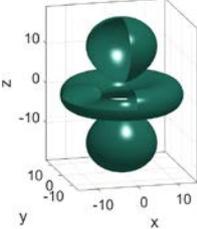 | 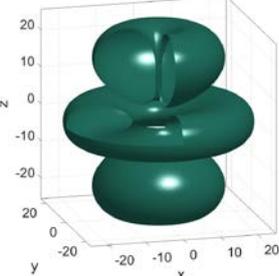 | 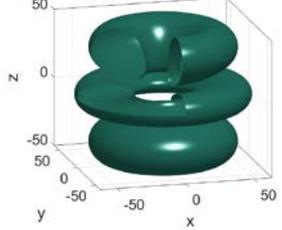 |
| 3 | 2 | 1 | 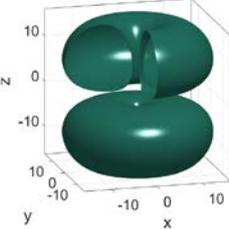 | 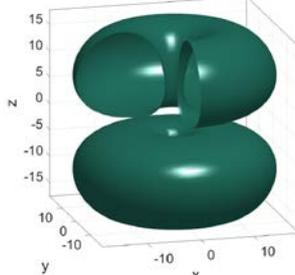 | 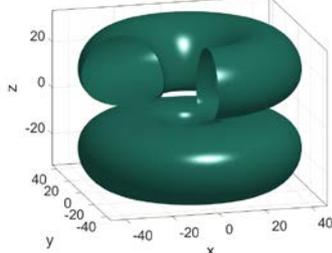 |
| 3 | 2 | 2 | 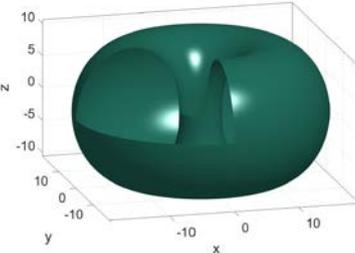 | 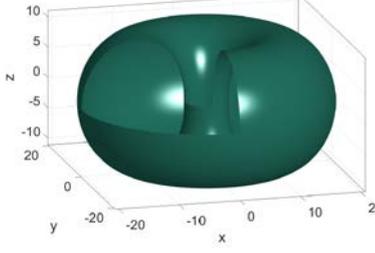 | 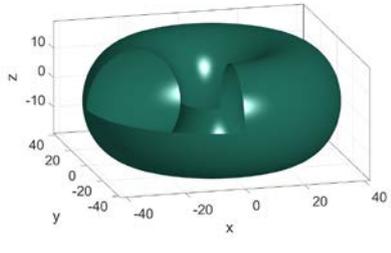 |
| 4 | 0 | 0 | 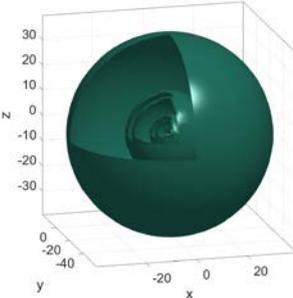 | 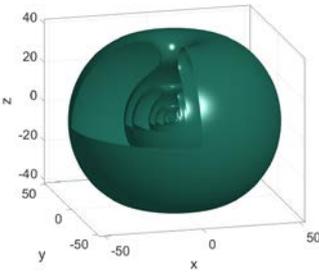 | 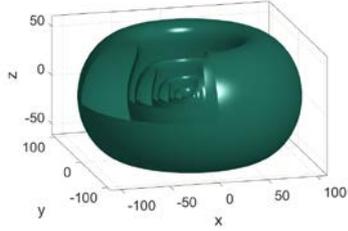 |

| 4 | 1 | 0 | 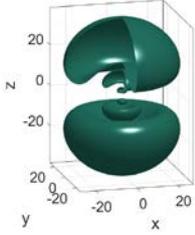 | 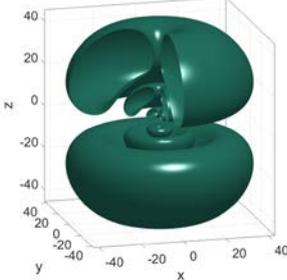 | 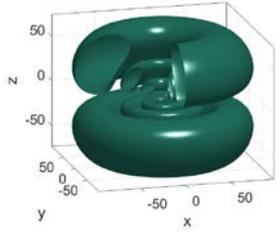 |
|---|---|---|---|---|---|
| 4 | 1 | 1 | 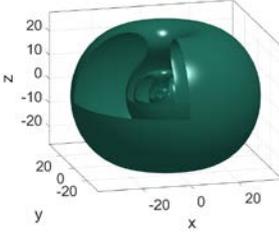 | 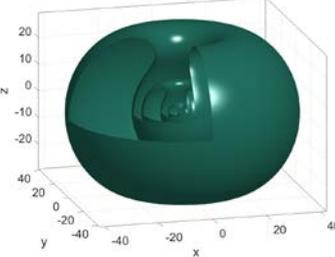 | 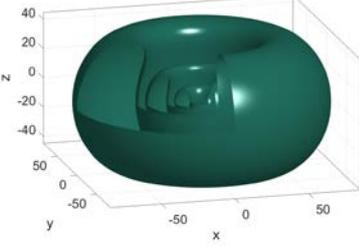 |
| 4 | 2 | 0 | 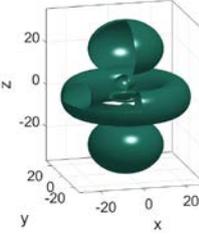 | 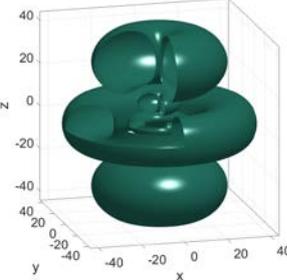 | 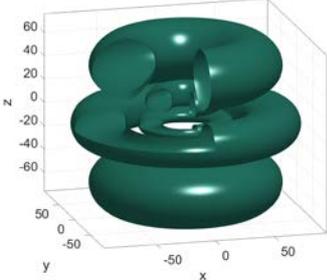 |
| 4 | 2 | 1 | 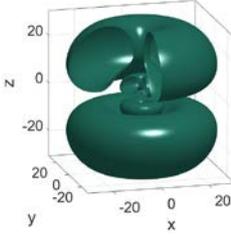 | 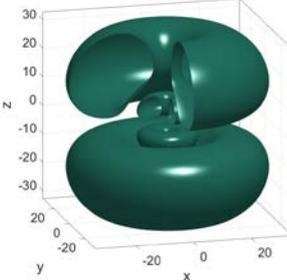 | 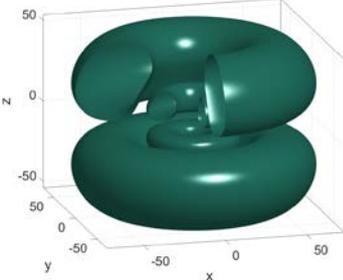 |
| 4 | 2 | 2 | 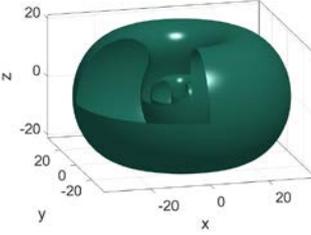 | 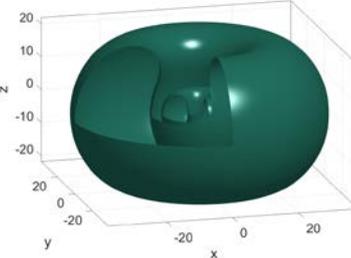 | 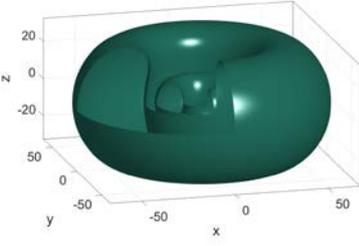 |
| 4 | 3 | 0 | 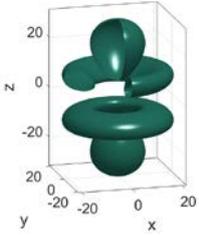 | 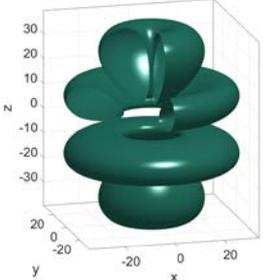 | 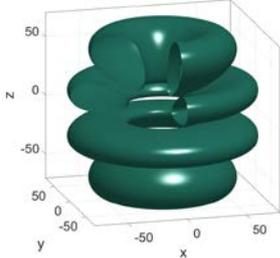 |

| 4 | 3 | 1 | 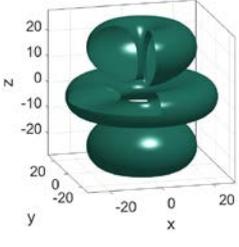 | 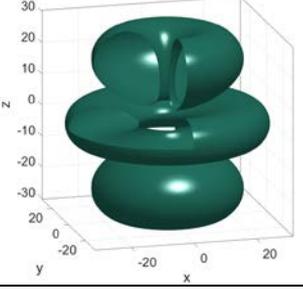 | 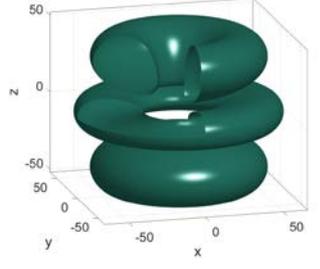 |
|---|---|---|---|---|---|
| 4 | 3 | 2 | 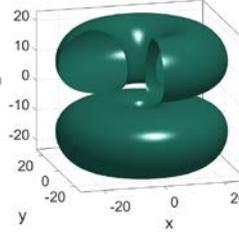 | 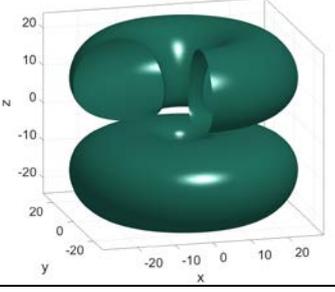 | 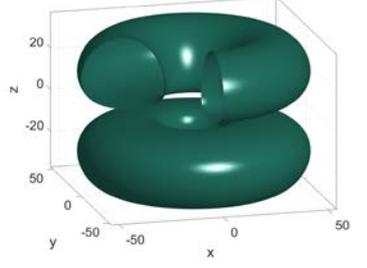 |
| 4 | 3 | 3 | 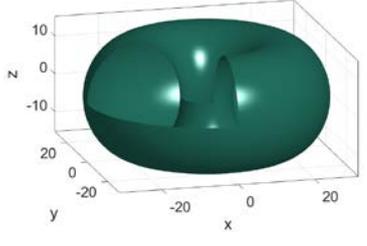 | 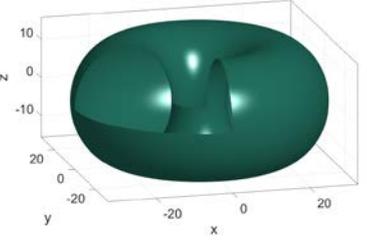 | 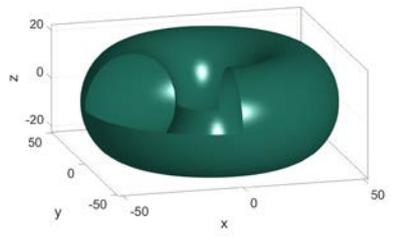 |
| 5 | 0 | 0 | 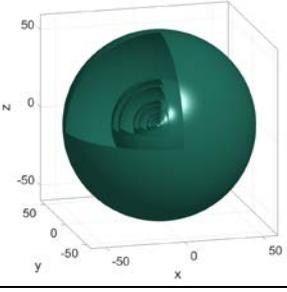 | 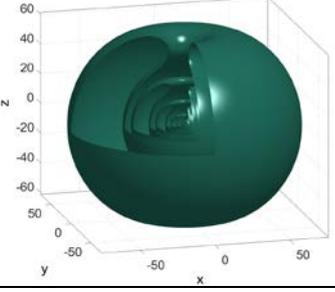 | 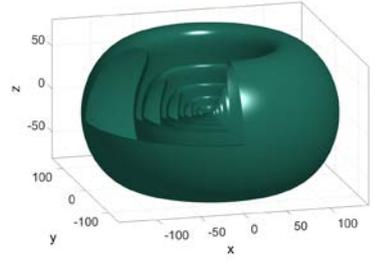 |
| 5 | 1 | 0 | 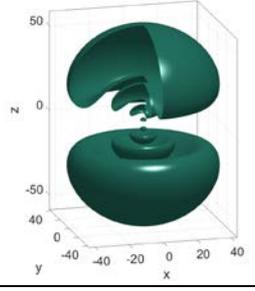 | 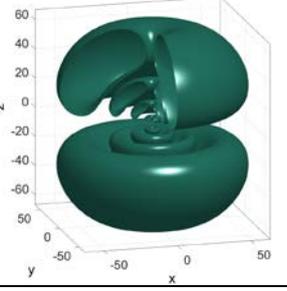 | 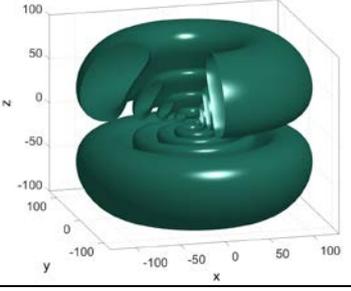 |
| 5 | 1 | 1 | 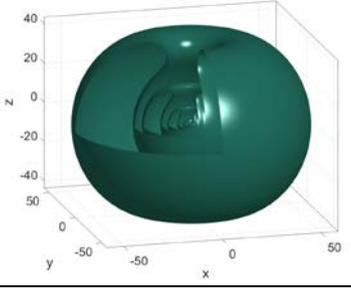 | 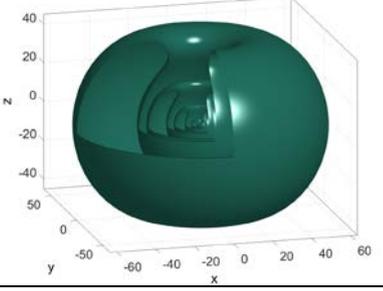 | 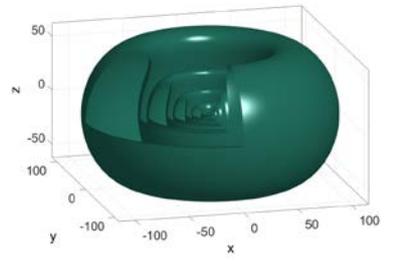 |

| 5 | 2 | 0 | 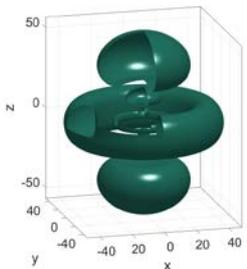 | 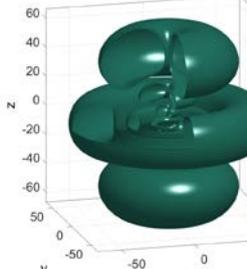 | 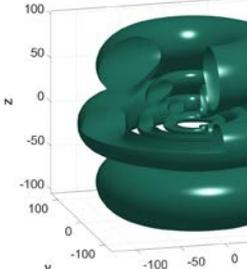 |
|---|---|---|---|---|---|
| 5 | 2 | 1 | 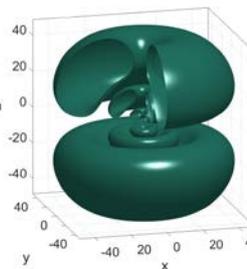 | 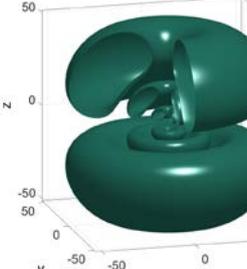 | 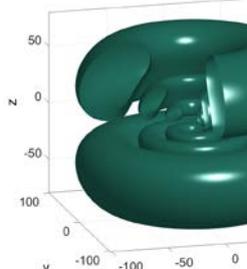 |
| 5 | 2 | 2 | 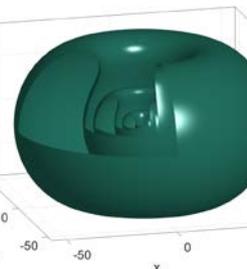 | 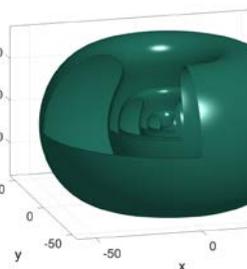 | 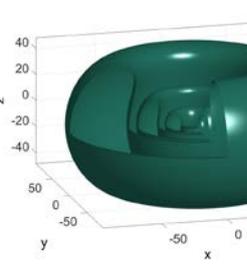 |
| 5 | 3 | 0 | 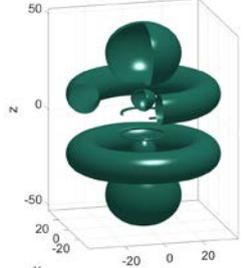 | 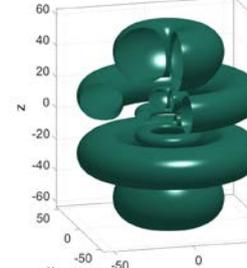 | 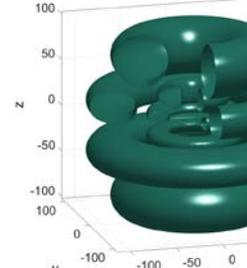 |
| 5 | 3 | 1 | 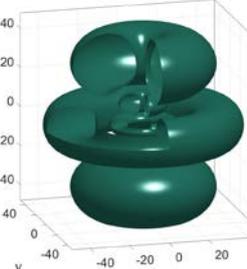 | 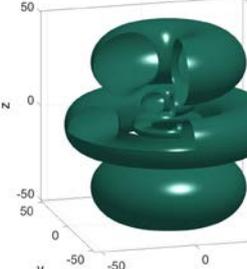 | 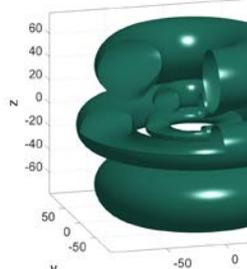 |
| 5 | 3 | 2 | 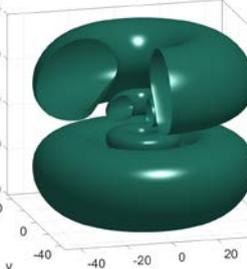 | 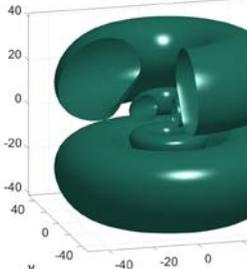 | 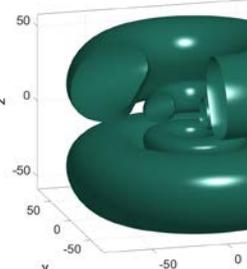 |

| 5 | 3 | 3 | 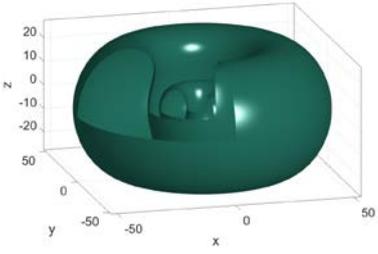 | 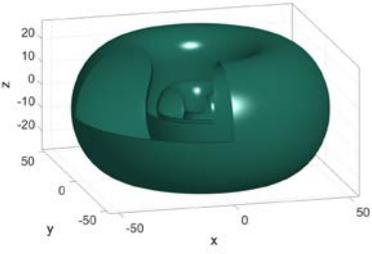 | 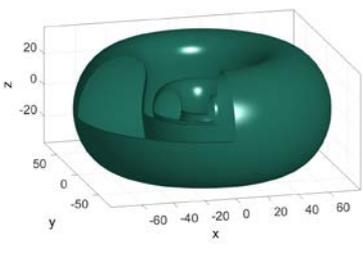 |
|---|---|---|---|---|---|
| 5 | 4 | 0 | 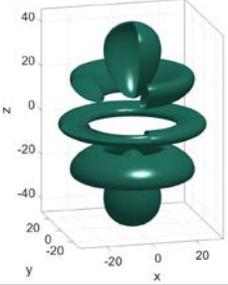 | 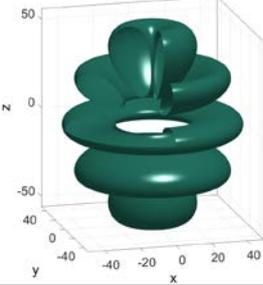 | 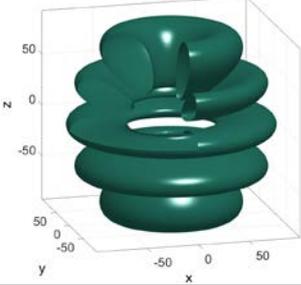 |
| 5 | 4 | 1 | 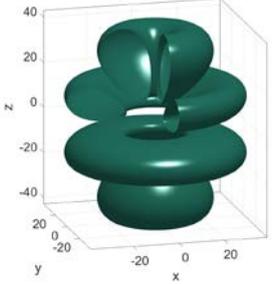 | 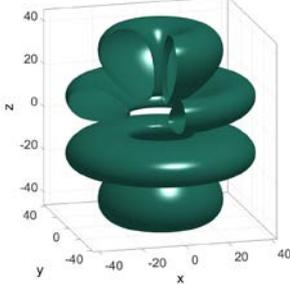 | 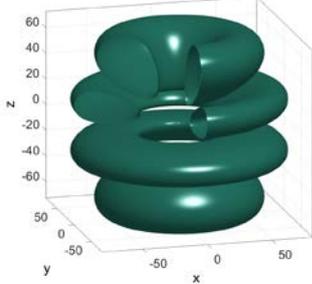 |
| 5 | 4 | 2 | 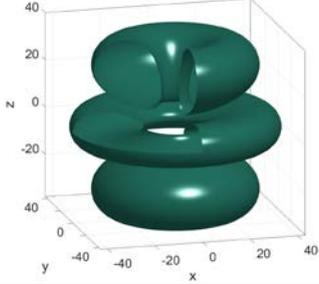 | 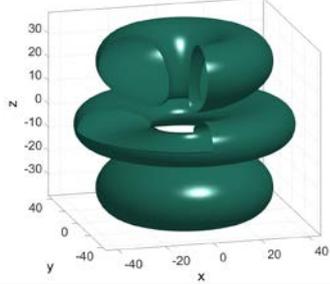 | 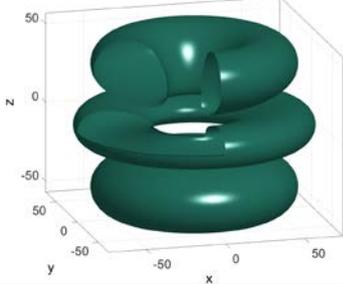 |
| 5 | 4 | 3 | 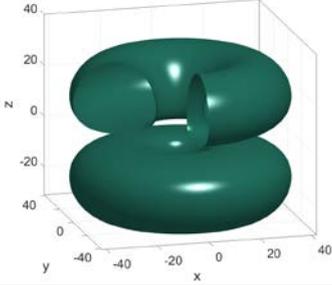 | 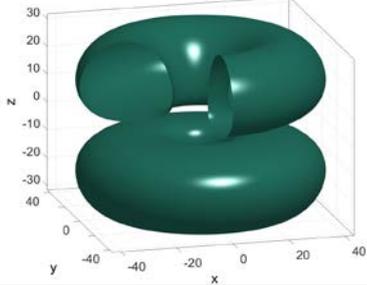 | 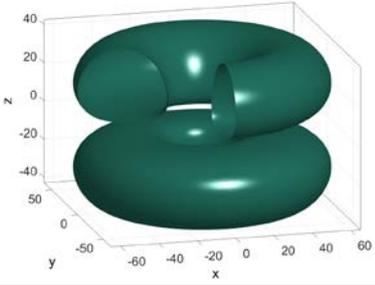 |
| 5 | 4 | 4 | 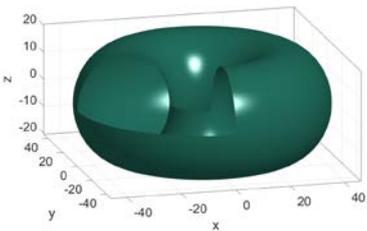 | 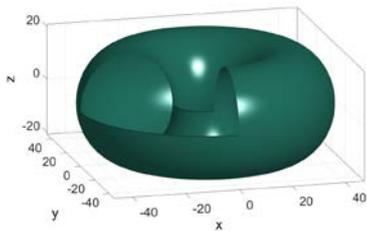 | 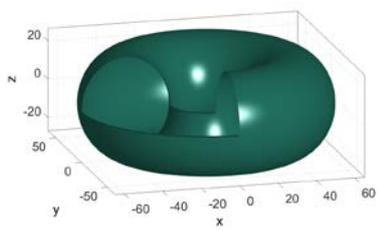 |

Table 3: Contour of the space probability distribution in two dimensional plane *yoz*

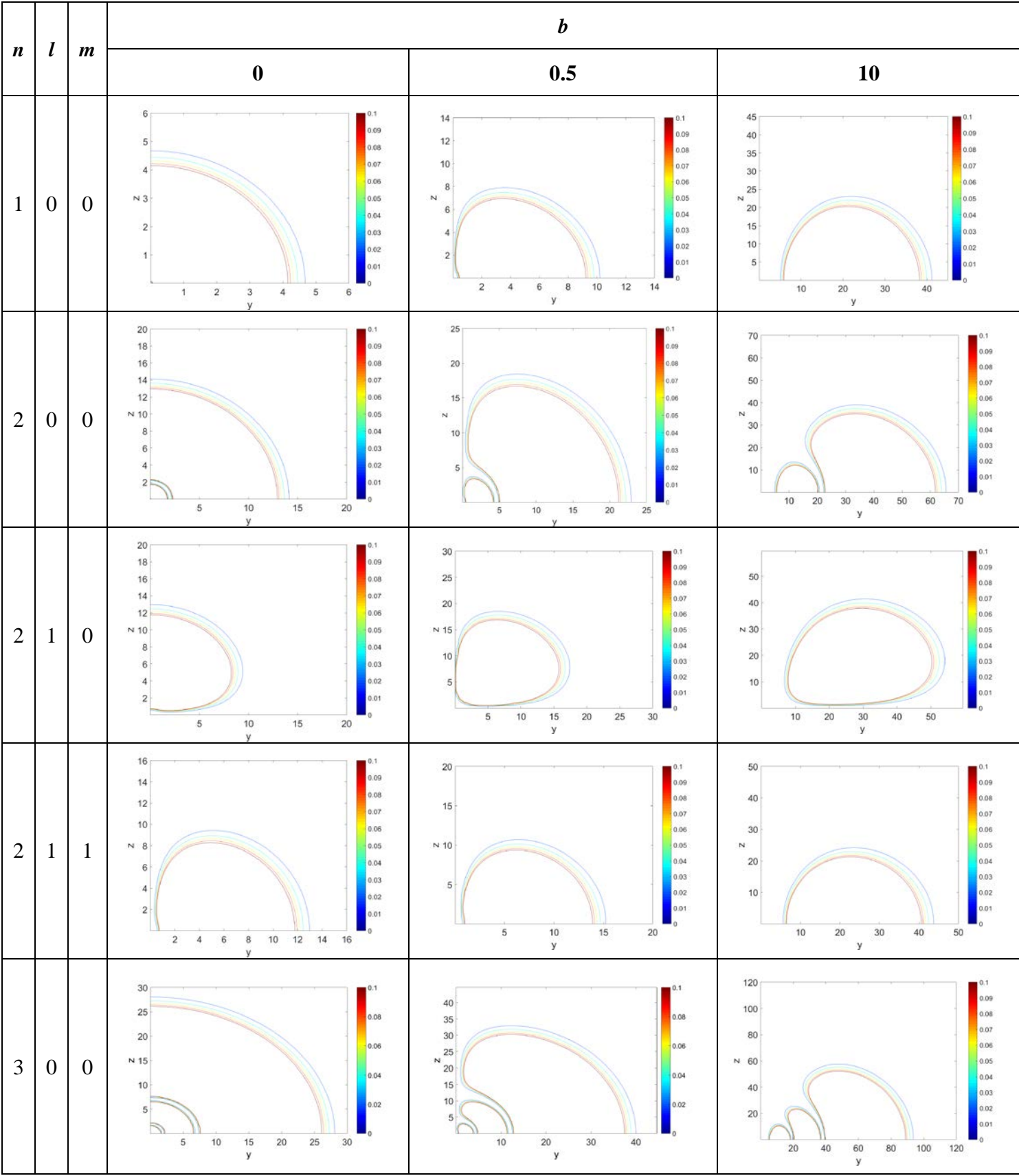

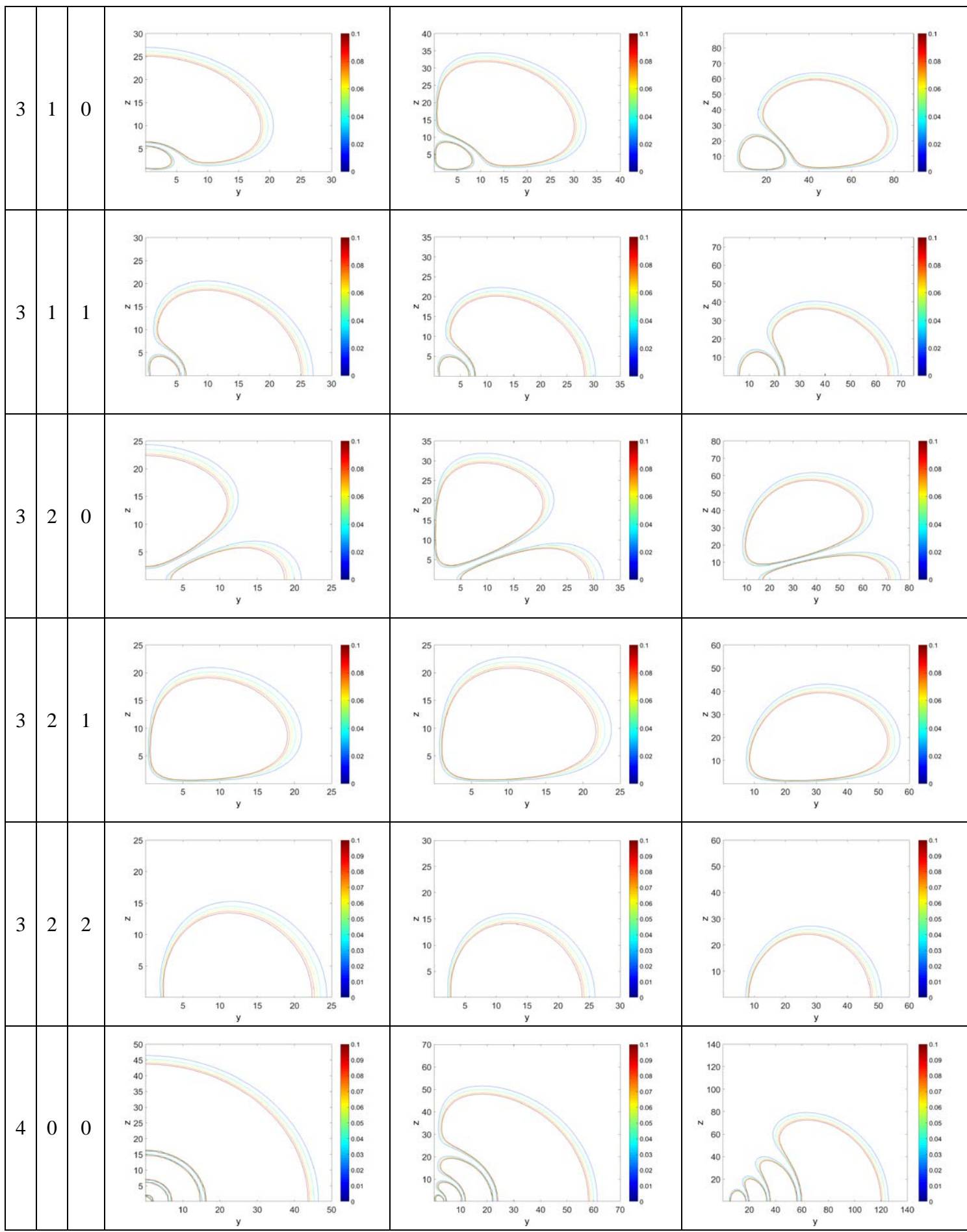

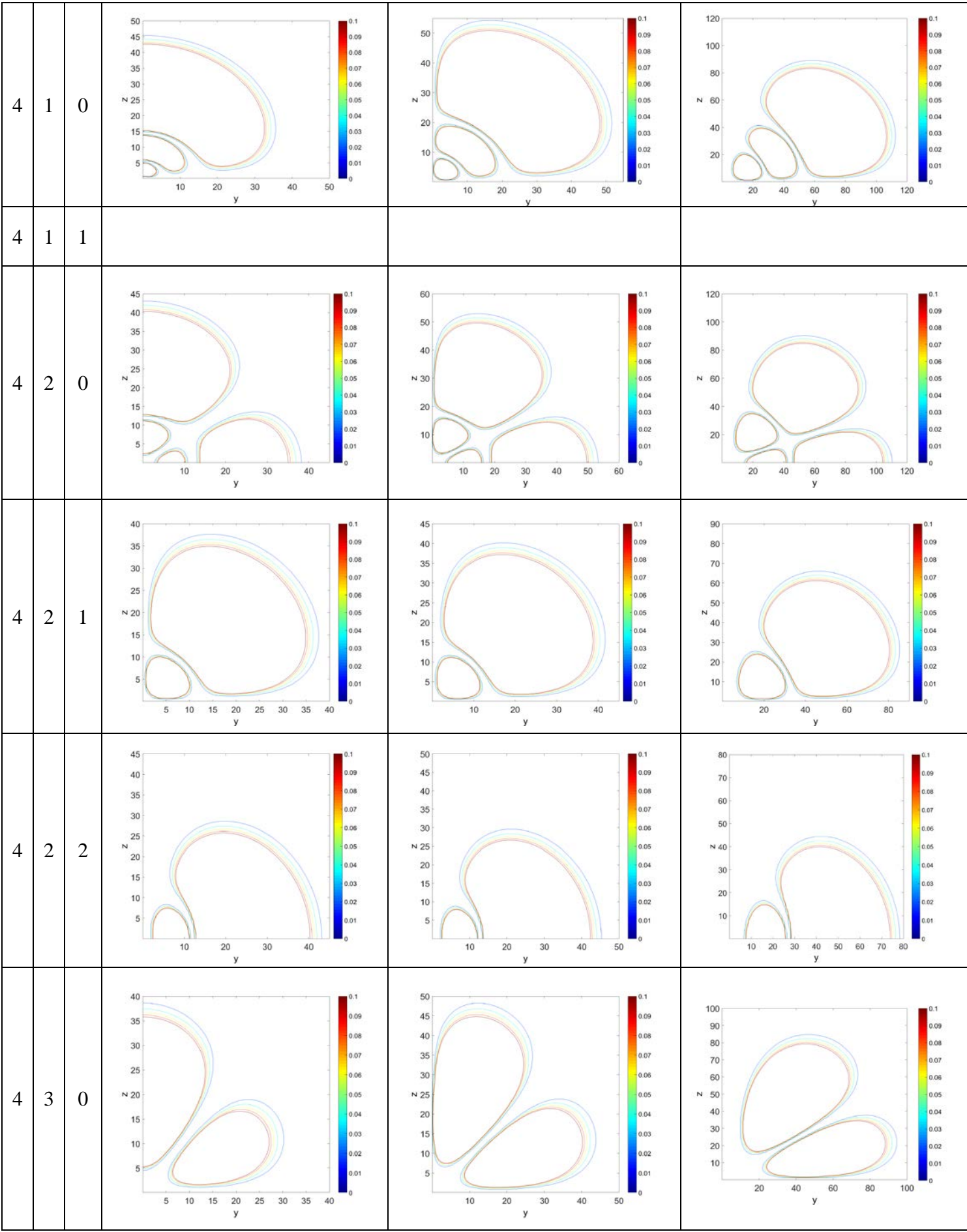

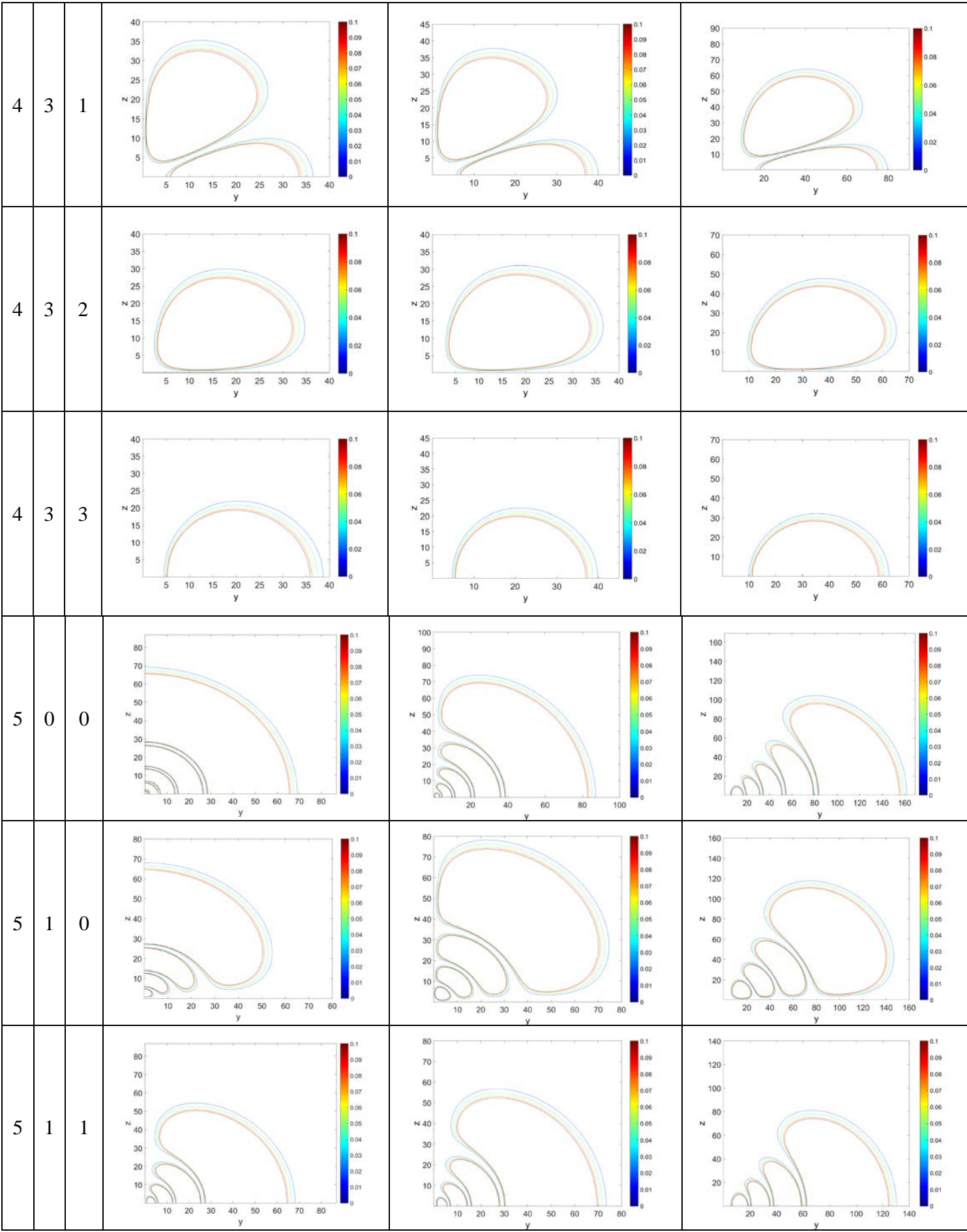

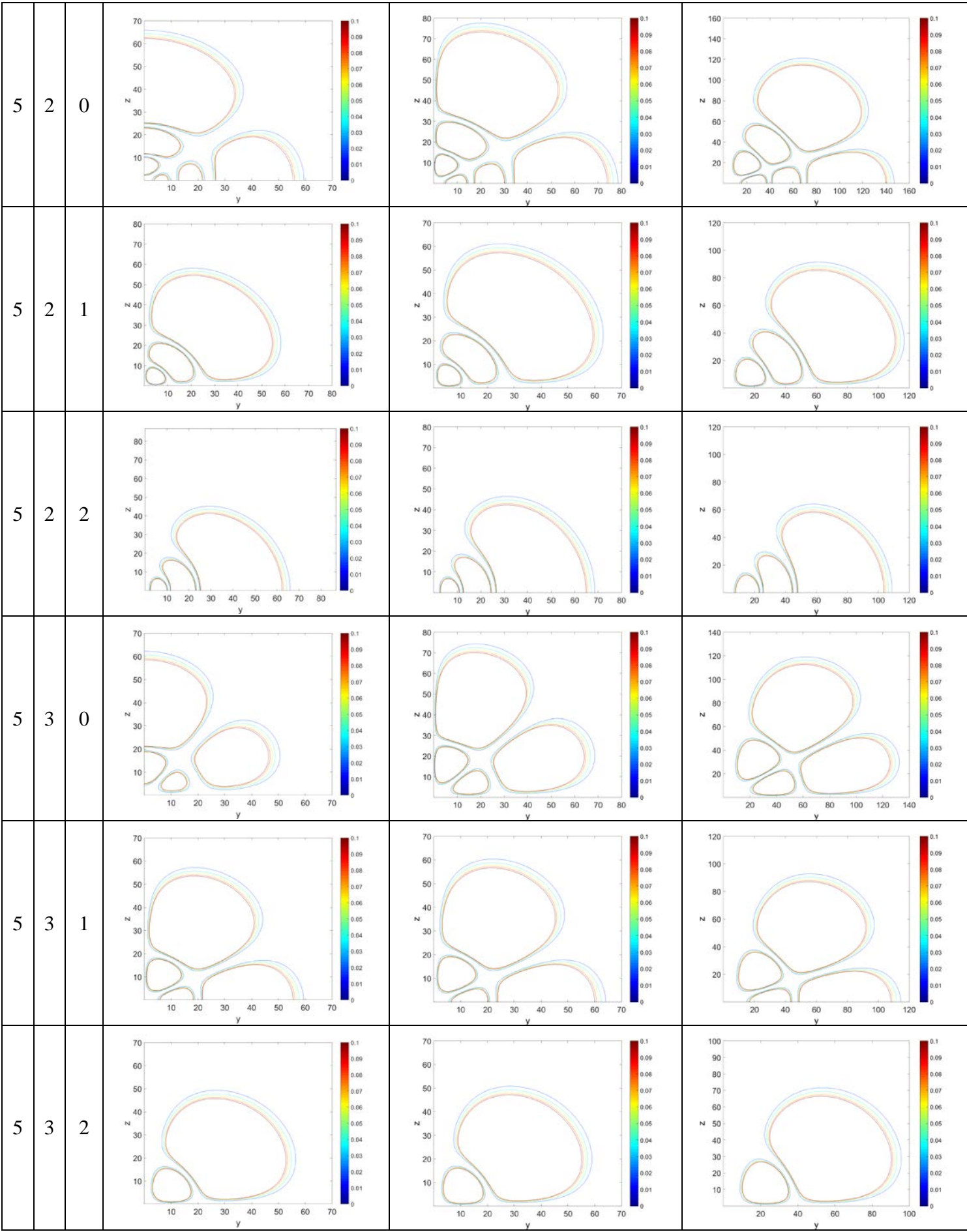

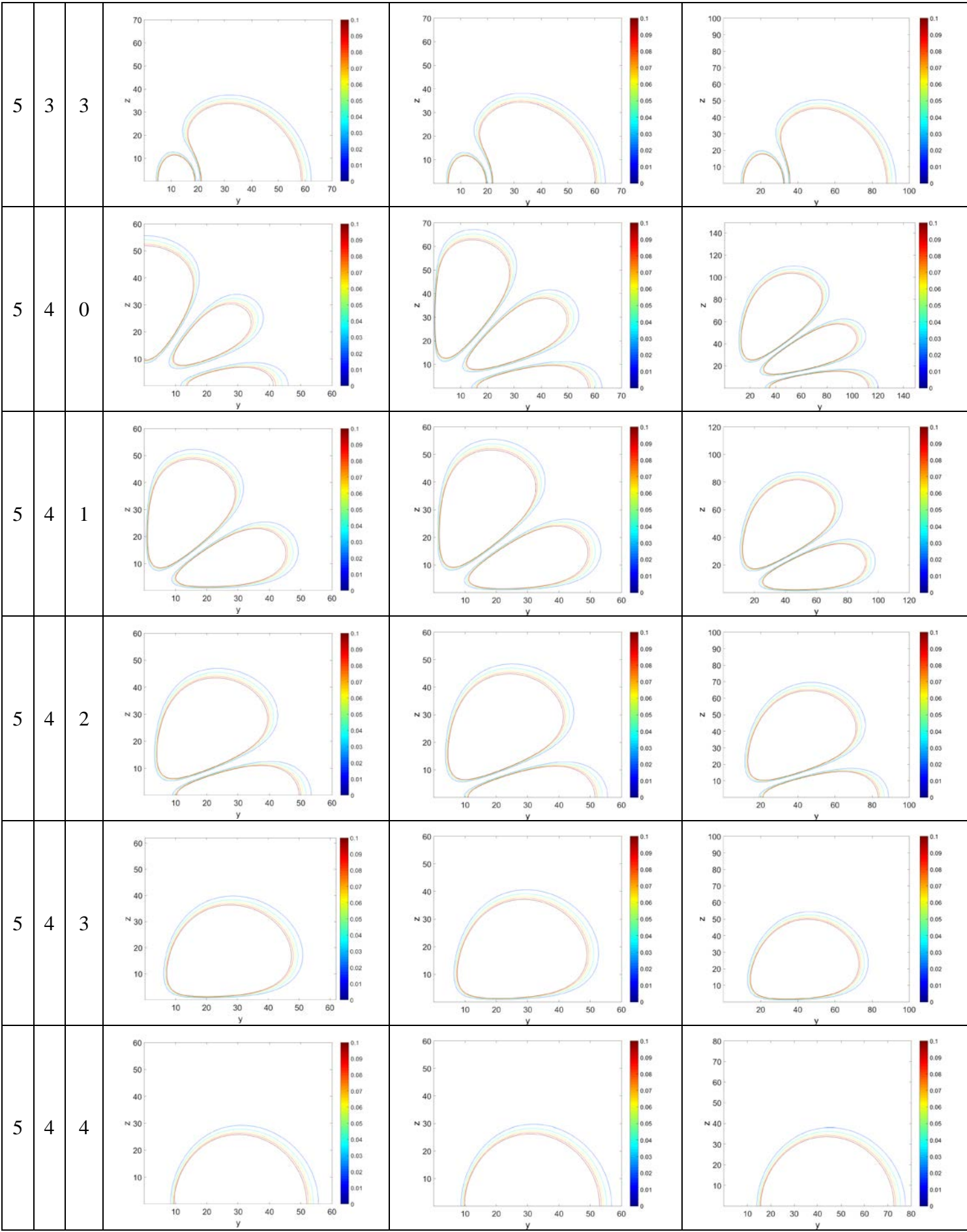

Table 4: Space probability distributions for different chosen relative probability values *P* for

(*n*, *l*, *m*)=(6, 5, 1)

| *P* | *b* | | |
|---|---|---|---|
| | 0 | 0.5 | 10 |
| 1% | 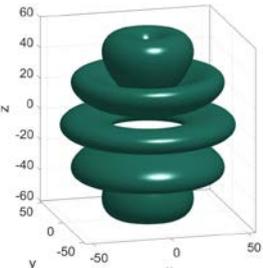 | 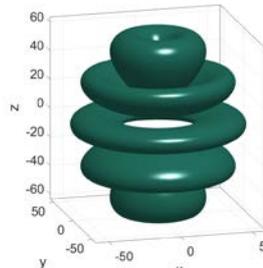 | 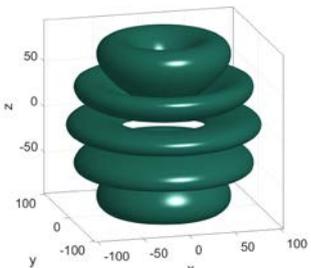 |
| 5% | 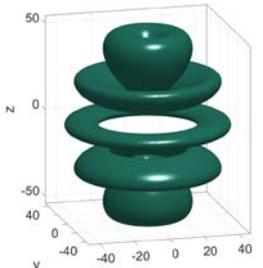 | 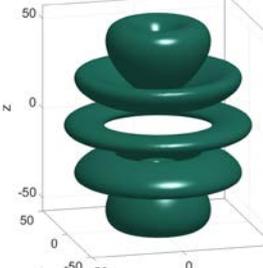 | 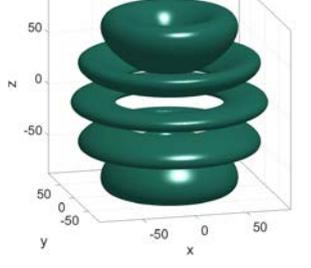 |
| 10% | 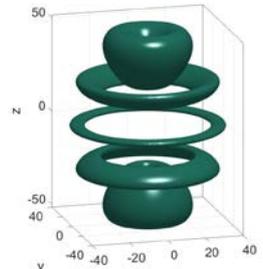 | 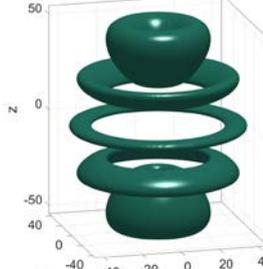 | 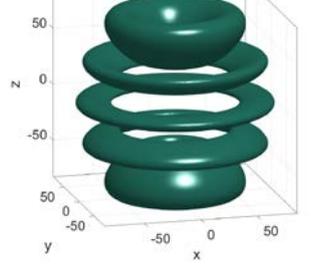 |
| 15% | 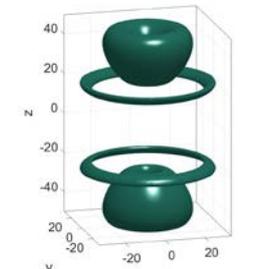 | 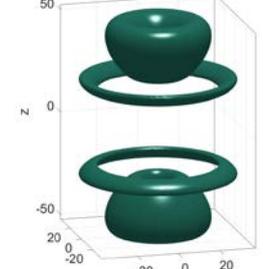 | 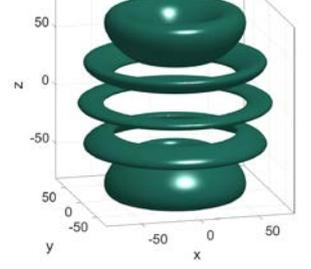 |
| 20% | 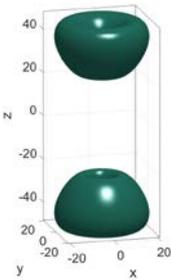 | 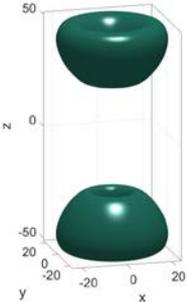 | 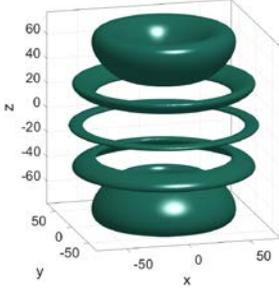 |
| 25% | 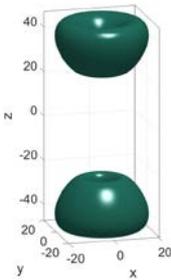 | 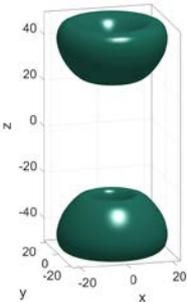 | 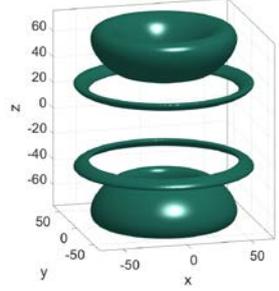 |
| 50% | 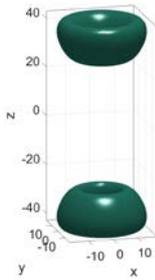 | 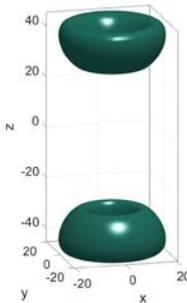 | 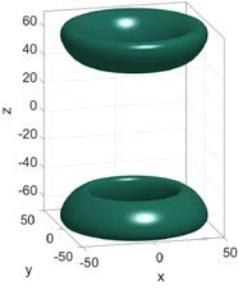 |
| 80% | 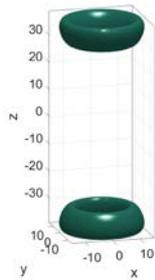 | 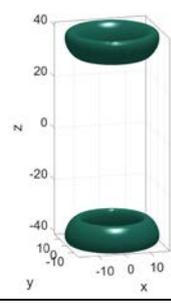 | 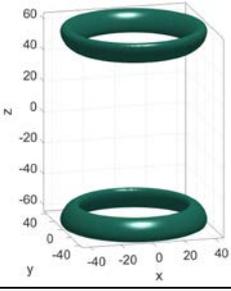 |
| 90% | 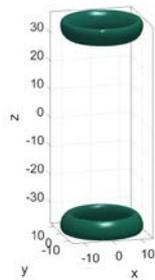 | 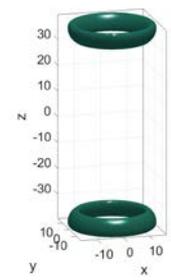 | 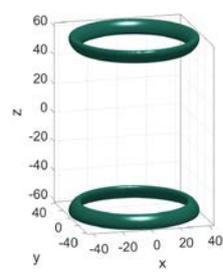 |
| 95% | 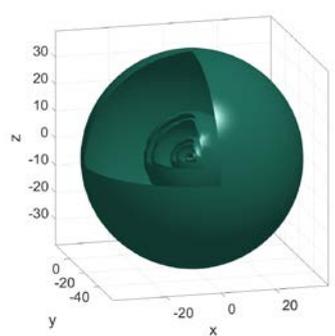 | 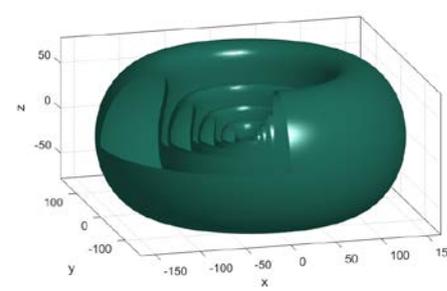 | 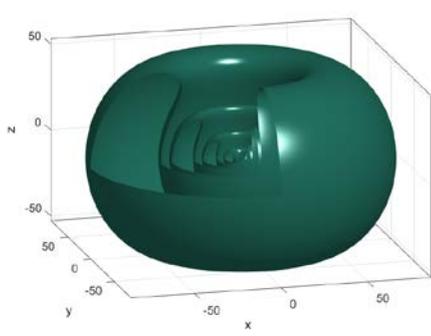 |

Table 5: Three dimensional isosurface representation of (4, 0, 0) for different positive values *b*

| *b* | Isosurface Representation | *b* | Isosurface Representation |
|---|---|---|---|
| **0** | 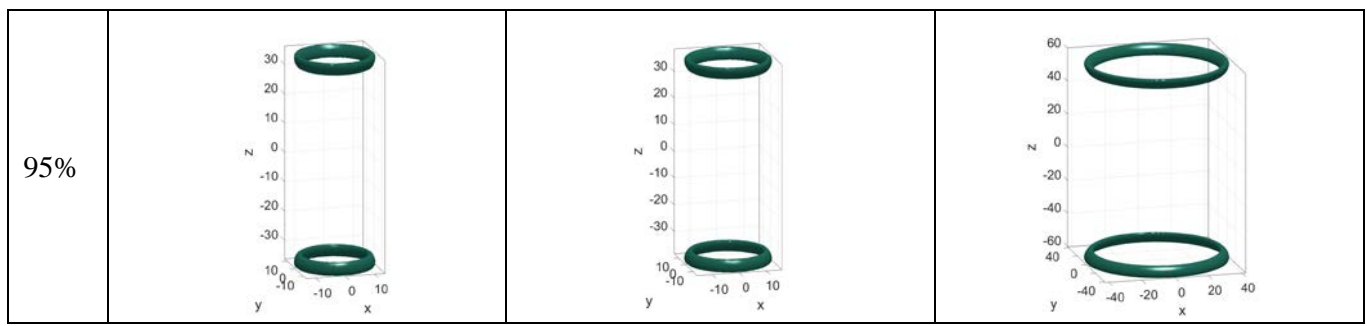 | **25** | |
| **5** | | **40** | |
| **10** | | **80** | |

Table 6: Space probability distributions for positive and negative parameter *b* for (5, 2, 1)

| *b* | Isosurface Representation | Contour Representation |
|---|---|---|
| 0.5 | 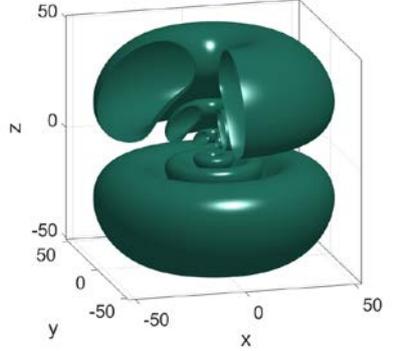 | 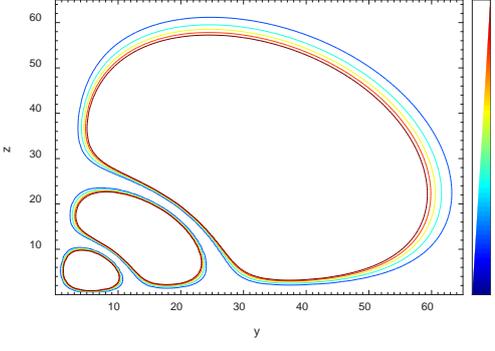 |
| 0 | 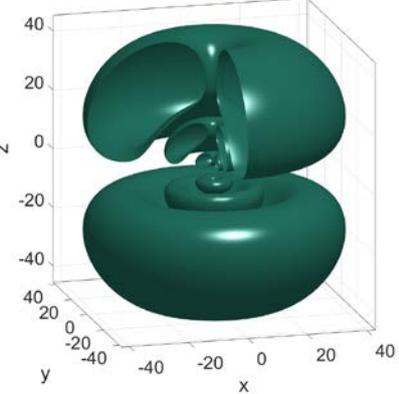 | 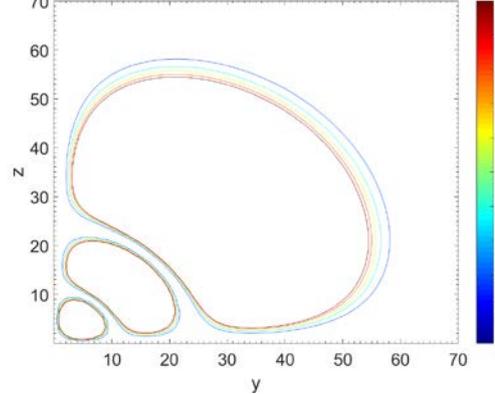 |
| -0.5 | 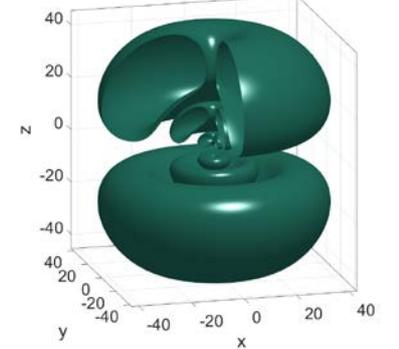 | 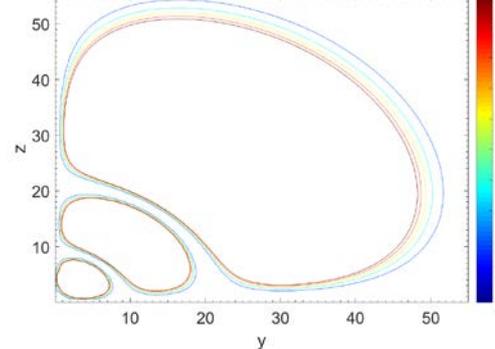 |

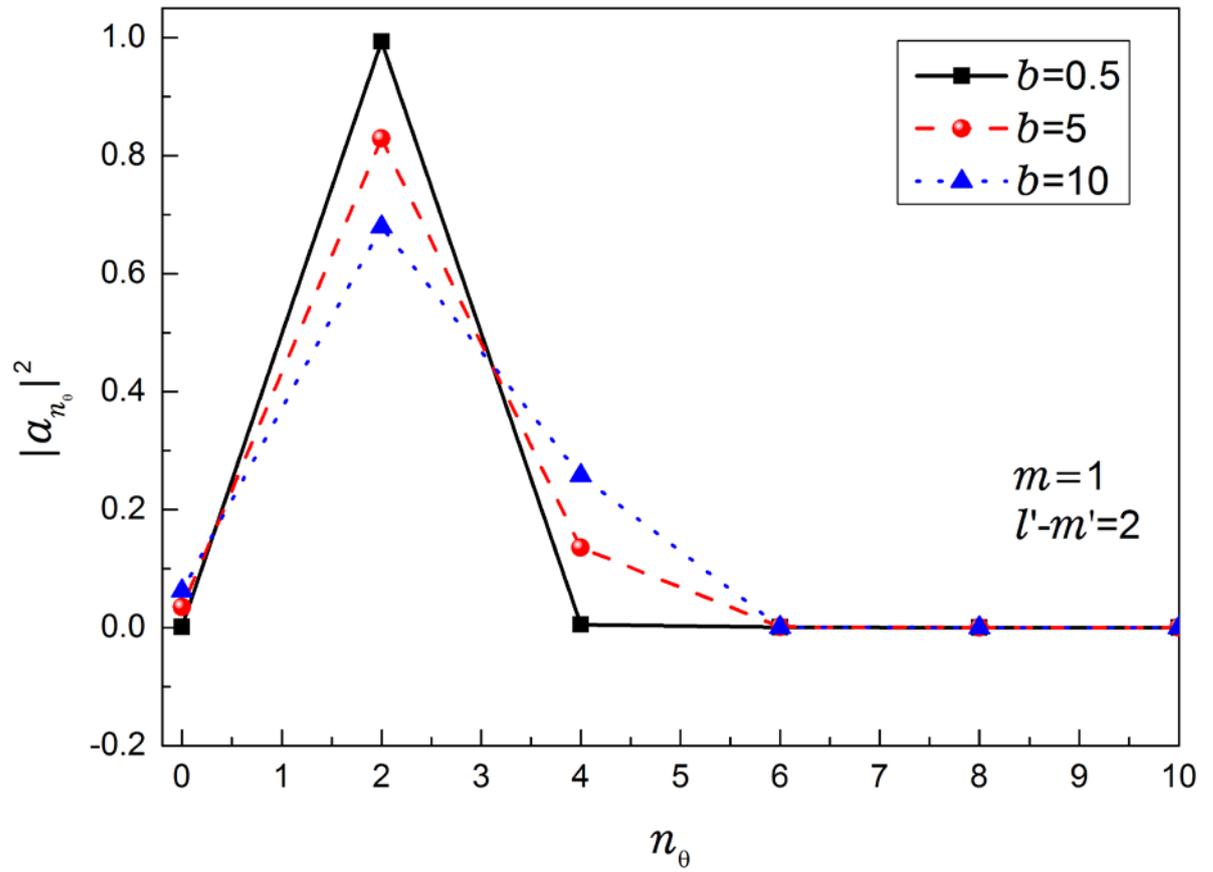

Fig. 1: Relation between $n_\theta$ and $|a_{n_\theta}|^2$ for different values $b$